# Microservices and Real-Time Processing in Retail IT: A Review of Open-Source Toolchains and Deployment Strategies

Aaditaa Vashisht
*Department of Information Science and Engineering*
*RV College of Engineering*
*Bangalore, India*
aaditaav.is21@rvce.edu.in

Mrs. Rekha B S (*Assistant Professor*)
*Department of Information Science and Engineering*
*RV College of Engineering*
*Bangalore, India*
rekhabs@rvce.edu.in

*Abstract*— **With the rapid pace of digital transformation, the retail industry is increasingly depending on real-time, scalable, and resilient systems to manage financial transactions, analyze customer behavior, and streamline order processing. This literature review explores how modern event-driven and microservices-based architectures, particularly those leveraging Apache Kafka, Spring Boot, MongoDB, and Kubernetes are transforming retail and financial systems. By systematically reviewing academic publications, technical white papers, and industry reports from recent years, this study synthesizes key themes and implementation strategies. The analysis reveals that technologies like Kafka and Spring Boot are instrumental in building low-latency, event-driven applications that support real-time analytics and fraud detection, while MongoDB, when deployed on Kubernetes, ensures fault tolerance and high availability in inventory and transaction systems. Kubernetes itself plays a crucial role in automating deployment and scaling of microservices. These findings provide valuable insights for industry practitioners aiming to design scalable infrastructures, identify research opportunities in hybrid deployment models, and offer educators a foundation to integrate modern system architectures into professional and technical communication training.**

*Keywords*— *Event-Driven Architecture, Apache Kafka, Spring Boot, MongoDB, Kubernetes, Microservices, Real-Time Analytics, Retail Technology, Financial Systems*

## I. Introduction

Retail and financial services are undergoing significant transformation with the rise of microservices, cloud-native platforms, and real-time data processing systems. As consumer demands for instant and personalized experiences continue to grow, businesses are increasingly turning to scalable and fault-tolerant infrastructures to stay competitive. Technologies like Apache Kafka, Spring Boot, MongoDB, and Kubernetes are playing a crucial role in powering real-time transaction handling, detecting fraud, managing inventory efficiently, and delivering personalized customer experiences. Despite their growing adoption, the fragmented nature of implementations and the scarcity of cross-domain comparative studies pose challenges for both practitioners and researchers. This integrative literature review addresses these gaps by examining how these technologies are practically applied to enhance scalability, fault tolerance, and responsiveness in retail and financial systems. Guided by the central research question, how Kafka, MongoDB, Spring Boot, and Kubernetes contribute to the development of resilient and real-time applications, this paper systematically analyzes academic and technical sources. It is organized into five key sections: a detailed methodology, a discussion of results, thematic synthesis, and a conclusion that outlines future directions for research and practical implementation.

## II. Research Methodology

This study adopts an integrative literature review methodology to synthesize insights from both academic research and real-world industry practices regarding the application of modern technologies, Apache Kafka, Spring Boot, MongoDB, and Kubernetes, in the retail and financial sectors. This approach was selected to capture the full scope of evolving cloud-native architectures, combining scholarly research with practitioner-oriented evidence. Rather than simply summarizing existing literature, the review aims to identify common themes, architectural patterns, performance characteristics, and deployment challenges across real-time retail and financial implementations. Inclusion criteria were strictly applied to select only publications from 2022 to 2024 that were peer-reviewed or industry-authored, featured one or more of the core technologies, and focused on real-time retail or financial applications. Conceptual-only papers, non-English sources, and those unrelated to the targeted domains were excluded. After initial screening, six high-quality sources, four academic papers and two technical white papers, were selected for in-depth analysis. These were annotated in Notion and mapped in an Excel matrix for metadata tracking.

Thematic coding was used to extract key concepts such as scalability under load, real-time event streaming, fault tolerance, integration ease, and data-intensive use case support. The analysis revealed consistent patterns in how these technologies are being deployed, Kafka and Spring Boot for real-time microservices, MongoDB with Kubernetes for high-volume data handling, and Kubernetes for resilient, cloud-native retail infrastructures. A qualitative comparative framework was applied to group studies by technological focus, allowing the review to draw evidence-based conclusions on the most effective tool combinations for various retail and financial applications.

III. Literature Review

Event-Driven Architecture: Harnessing Kafka and Spring Boot for Scalable, Real-Time Applications: This publication explores how event-driven architecture (EDA) can be implemented using Apache Kafka and Spring Boot to support scalable and responsive systems. It emphasizes the role of Kafka in enabling asynchronous communication between services and highlights Spring Boot's simplicity in developing lightweight, microservice-based applications. The paper helped me better understand how combining Kafka and Spring Boot can create efficient, real-time processing systems that are well-suited for modern enterprise environments.[1]

Real-time Event Joining in Practice With Kafka and Flink: This research focuses on practical strategies for joining multiple real-time data streams using Apache Kafka and Apache Flink. It discusses challenges in synchronizing data across streams and demonstrates how window-based joins and stateful stream processing in Flink can handle complex use cases like fraud detection or behavioral analytics. This study provided valuable insight into advanced stream processing techniques and the importance of real-time correlation in event-driven systems.[2]

Real-time Data Streaming: Exploring Real-Time Data Streaming with Tools like Apache Kafka: This whitepaper outlines the fundamental concepts behind real-time data streaming, with Apache Kafka as the central messaging system. It explains Kafka's architecture, including topic partitioning, message durability, and stream persistence, in the context of large-scale data processing. The document also presents real-world applications across various industries. This resource helped clarify how Kafka supports continuous data flows and underpins responsive digital systems.[3]

Deploying MongoDB Clusters on Kubernetes for High-Volume Retail Transactions : In this technical whitepaper, MongoDB Inc. explains how deploying MongoDB on Kubernetes enhances scalability, resilience, and performance in high-volume retail applications. It covers operational aspects like automated failover, sharding, and cluster management. This work helped me understand how container orchestration through Kubernetes simplifies the deployment and scaling of NoSQL databases in retail environments where data volume and uptime are critical.[4]

Scalable Microservices Deployment Using Kubernetes in Retail Environments: This conference paper presents a detailed approach to deploying microservices using Kubernetes within the retail sector. It highlights how Kubernetes manages service discovery, load balancing, and automated scaling to support modular application design. The paper also explores the integration of DevOps practices such as CI/CD pipelines. This work provided insights into building flexible, fault-tolerant systems that can adapt to the dynamic needs of the retail industry.[5]

Leveraging MongoDB for Real-Time Inventory Management in E-Commerce: This study investigates the use of MongoDB for managing inventory in e-commerce platforms. It explores how features like change streams and schema flexibility allow for real-time updates and tracking. By integrating MongoDB into backend systems, businesses can monitor stock levels accurately and respond quickly to changes in demand. This paper reinforced my understanding of how NoSQL databases support agile and efficient inventory management in digital commerce.[6]

IV. Results and Discussions

The sources were analyzed based on their technological focus, application within the retail or financial domain, and the core benefits and challenges identified. A comparative synthesis is presented below, highlighting the patterns and divergences in their findings.

TABLE I
Comparative Analysis of Reviews Literature

| Title | Technologies | Use Case | Key Contributions |
|---|---|---|---|
| *Event-Driven Architecture: Harnessing Kafka and Spring Boot for Scalable, Real-Time Applications* | Apache Kafka, Spring Boot | Financial transactions and customer order processing | Demonstrated how Kafka's pub-sub model and Spring Boot's microservice support create a scalable, low-latency system ideal for real-time financial operations. |

| *Real-time Event Joining in Practice With Kafka and Flink* | Apache Kafka, Apache Flink | Customer behavior tracking and recommendation engines | Emphasized real-time stream joining and personalization pipelines, showing how event-driven models help with continuous machine learning and dynamic user experience adaptation. |
|---|---|---|---|
| *Real-time Data Streaming: Exploring real-time data streaming with tools like Apache Kafka* | Apache Kafka, Spring Boot | Inventory alerts, fraud detection, dynamic pricing | Highlighted the real-world impact of integrating Kafka with Spring Boot for reliable alert systems and time-sensitive decision-making in retail. |
| *Deploying MongoDB Clusters on Kubernetes for High-Volume Retail Transactions* | MongoDB, Kubernetes | Order logs, receipts, high-volume retail transactions | Provided deployment best practices for combining MongoDB's NoSQL performance with Kubernetes' auto-scaling and failover capabilities in retail transaction processing. |
| *Scalable Microservices Deployment Using Kubernetes in Retail Environments* | Kubernetes | Product catalog and order management | Validated Kubernetes' role in deployment automation, zero-downtime scaling, and seasonal workload management in retail microservices infrastructure. |
| *Leveraging MongoDB for Real-Time Inventory Management in E-Commerce* | MongoDB | Real-time inventory updates | Illustrated the schema-flexibility of MongoDB and its suitability for fast-changing e-commerce environments where agility and performance are critical. |

A. Thematic Synthesis

The literatures studied reveal 3 core themes common across all papers:

Scalability and Performance under Load: All sources emphasized the necessity of systems that can scale both vertically and horizontally. Kubernetes and Kafka played pivotal roles in ensuring elasticity, high throughput, and low latency, especially during retail spikes or financial transaction bursts.

Real-Time Data Processing: Kafka and Flink emerged as primary enablers of real-time streaming and event processing. Papers by Saket et al. and Venkatachalapathi clearly demonstrated how these tools enabled immediate customer response, fraud detection, and personalization—driving improved user experiences.

Modularity and Integration: Spring Boot and MongoDB were repeatedly praised for their ease of integration into microservice ecosystems. MongoDB's schema-less design provided flexibility for rapidly evolving product catalogs and pricing systems, while Spring Boot reduced boilerplate code and streamlined service deployment.

B. Technological Trade-Offs

The analysis also revealed several important technological trade-offs that organizations must consider when adopting these tools. Apache Kafka and Apache Flink provide powerful capabilities for real-time event streaming and analytics, but they demand considerable infrastructure and specialized expertise to operate effectively, which can raise the barrier to entry. MongoDB offers flexibility and high performance across a variety of workloads; however, it may fall short in delivering the strict consistency guarantees often required in financial systems, necessitating the implementation of additional architectural safeguards. Kubernetes enhances system robustness and scalability, but its adoption introduces operational complexity, particularly around service discovery, container orchestration, and system monitoring—making seamless DevOps integration a critical requirement for successful deployment.

*C. Identified Research Gaps*

Despite notable technological progress discussed in the reviewed literature, several research gaps remain that constrain a comprehensive understanding of how these systems perform in real-world, large-scale retail and financial environments. Firstly, while individual technologies like Apache Kafka, MongoDB, Spring Boot, and Kubernetes are well-documented, there is a noticeable lack of empirical studies that explore their integrated deployment within a unified architecture. The combined performance and trade-offs such as using Kafka alongside MongoDB under Kubernetes orchestration for real-time transactions have not been adequately evaluated. Secondly, existing studies largely focus on short-term implementation metrics and scalability, with minimal attention to long-term performance under evolving workloads, sustained peak traffic, or microservice churn, which are crucial for assessing resilience and total cost of ownership. Another key gap is the absence of standardized benchmarks across studies; variations in performance metrics like latency, availability, and integration effort hinder meaningful comparisons and replication. Additionally, security and regulatory compliance, which are critical for both retail and financial sectors, are underexplored. Mechanisms for secure data streaming in Kafka or data integrity controls in MongoDB are often only briefly mentioned, if at all. Furthermore, most research emphasizes technical implementations without linking them to measurable business outcomes such as improved customer retention, conversion rates, or operational efficiency. User experience, as a result, remains an overlooked dimension. Lastly, there is a lack of attention to small and medium enterprises (SMEs), as most case studies revolve around large-scale enterprises. Research on lightweight, cost-effective, and skill-accessible deployment models tailored for SMEs is notably absent, highlighting a need for more inclusive studies that reflect the broader retail technology landscape.

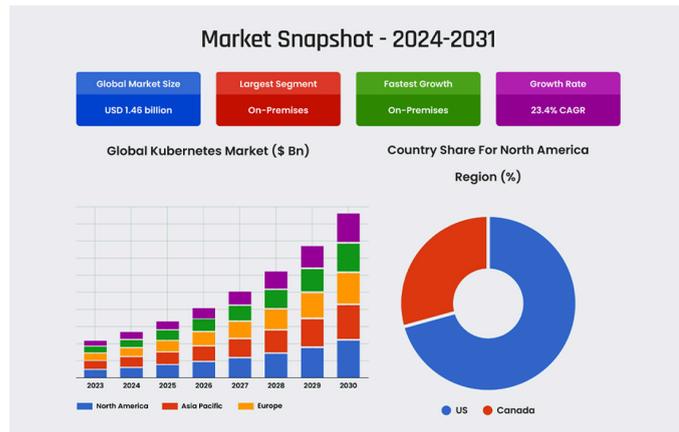

Fig. 1 Predicted market snapshot of kubernetes in 2024-2031 for USA by CNCF and Skyquest

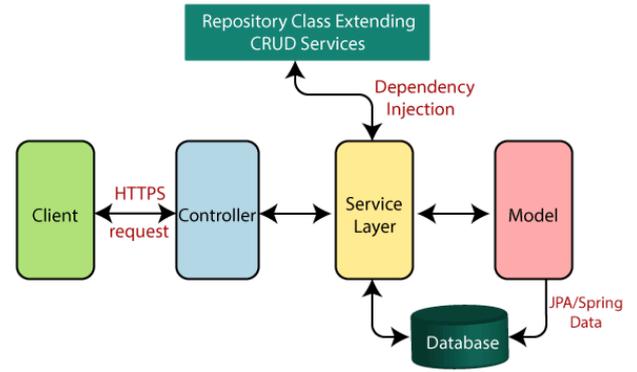

Fig. 2 Basic Spring Boot architecture for a system

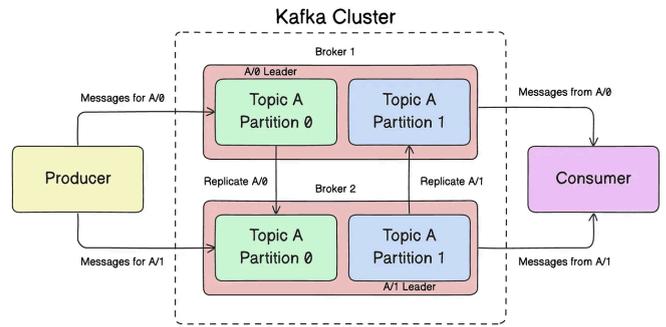

Fig. 3 The overall architecture of kafka

V. CONCLUSIONS AND FURTHER RESEARCH

This integrative literature review synthesizes findings from six key studies and technical whitepapers, revealing that the integration of Apache Kafka, Spring Boot, MongoDB, and Kubernetes provides a resilient and scalable foundation for modern retail and financial systems. These technologies collectively support critical requirements such as low-latency data processing, real-time customer interaction, flexible schema management, and fault-tolerant deployment infrastructures. By bridging the gap between empirical academic studies and industrial whitepapers, this review offers a consolidated understanding of how event-driven architectures (EDAs), container orchestration, and distributed NoSQL databases are being leveraged to meet dynamic market demands. The findings highlight the capacity of these tools to support diverse use cases, including real-time inventory tracking, fraud detection, customer behavior analysis, and transaction processing. This work contributes to the field by mapping the landscape of current best practices in data-intensive application architecture. It serves as a reference point for developers, architects, and IT managers seeking to align their system designs with proven, scalable, and industry-aligned technology stacks.

## D. Implications for Practice, Education & Research

For technology practitioners and system architects, this review reinforces the practical value of adopting microservices-based architecture, stream processing, and containerization in retail and financial domains. Kafka's ability to handle large volumes of streaming data, paired with Spring Boot's lightweight integration capabilities, offers a robust solution for developing loosely coupled, scalable systems. MongoDB provides the schema flexibility needed to accommodate ever-evolving product and customer data, while Kubernetes ensures system reliability and elasticity during peak traffic periods. Organizations aiming to modernize legacy systems or build cloud-native applications can utilize this integrated stack to achieve higher throughput, improve customer engagement through real-time responsiveness, and reduce system downtime.

In academic settings, these findings underscore the importance of integrating emerging cloud-native technologies into computer science and information systems curricula. Educators should incorporate modules on Kafka, MongoDB, Kubernetes, and Spring Boot into systems architecture and software engineering courses. Hands-on labs and case studies involving these tools will help students gain practical insights into their capabilities, limitations, and real-world applications, thus better preparing them for careers in technology-driven industries.

From a research standpoint, this review identifies several fruitful directions for further inquiry. While the reviewed literature demonstrates the efficacy of individual technologies and some integrated use cases, there is a notable lack of systematic comparisons, performance evaluations, and cross-platform studies. More empirical studies are needed to assess the impact of these architectures on system performance, user satisfaction, and cost-effectiveness across diverse deployment environments. Moreover, the evolving nature of cloud services and open-source tools calls for continuous monitoring and evaluation. As new alternatives to Kafka (e.g., Apache Pulsar, Red Panda) and MongoDB (e.g., Couchbase, Redis) gain traction, comparative analysis studies will be essential to determine optimal tool selection for specific workloads and operational goals.

## E. Limitations

While this review provides valuable insights, it is constrained by several limitations. First, the number of high-quality sources focused specifically on real-world retail and financial implementations is limited. Much of the literature is either conceptual or lacks long-term performance metrics and scalability benchmarks. Second, the review focuses primarily on studies published in the past years, potentially overlooking earlier foundational work or developments emerging in 2025 and beyond. Additionally, many of the technical whitepapers included are industry-authored and may reflect inherent promotional bias, lacking peer-reviewed rigor. This could influence the objectivity of some reported benefits, especially in regard to scalability and integration ease.

## F. Further Research Directions

For technology practitioners and system architects, this rTo overcome the identified limitations and drive further innovation, future research should delve into several key areas. First, comparative streaming analyses are needed to benchmark Apache Kafka against emerging alternatives like Apache Pulsar and Red Panda, focusing on metrics such as latency, throughput, fault tolerance, and ease of integration. Additionally, studies should explore the cost-performance trade-offs of deploying MongoDB clusters across leading cloud platforms AWS, Azure, and GCP particularly under burst traffic conditions common in e-commerce environments. Longitudinal case studies would also provide valuable insights by tracking the long-term impact of event-driven architectures in retail enterprises, evaluating not only system performance but also maintenance costs and customer experience improvements over time. Furthermore, research into hybrid architecture optimization such as the combined use of Kafka and Flink or MongoDB with Spring Boot on Kubernetes should investigate strategies for efficient resource allocation, load balancing, and orchestration performance under varying network conditions. Lastly, more customer-focused research is necessary to quantify how real-time processing systems influence business outcomes, including metrics like customer satisfaction, conversion rates, cart abandonment, and the effectiveness of personalization efforts.

## ACKNOWLEDGMENT

The author would like to express heartfelt gratitude to Mrs. Rekha B. S., guide and mentor at RV College of Engineering, for her invaluable guidance, encouragement, and constructive feedback throughout the course of this research project. Her expertise and support played a crucial role in shaping the direction and quality of this work. The author also acknowledges the faculty and staff of the Department of Information Science and Engineering at RV College of Engineering for providing a conducive academic environment and access to necessary resources.Special thanks are extended to the authors and organizations behind the technical papers, white papers, and research articles referenced in this review for their significant contributions to the field of event-driven and real-time architectures in retail and finance.